\documentclass[conference]{IEEEtran}
\IEEEoverridecommandlockouts
\usepackage{cite}
\usepackage{amsmath,amssymb,amsfonts}
\usepackage{algorithmic}
\usepackage{graphicx}
\usepackage{textcomp}
\usepackage{xcolor}
\def\BibTeX{{\rm B\kern-.05em{\sc i\kern-.025em b}\kern-.08em
    T\kern-.1667em\lower.7ex\hbox{E}\kern-.125emX}}

\usepackage{tikz}
\newcommand\copyrightnotice[1]{
	\begin{tikzpicture}[remember picture,overlay]
		\node[anchor=south,yshift=20pt] at (current page.south) {\fbox{\parbox{\dimexpr\textwidth-\fboxsep-\fboxrule\relax}{#1}}};
	\end{tikzpicture}
}

\usepackage{color}
\usepackage{etoolbox}
\usepackage{multirow}
\usepackage{siunitx}
\usepackage{balance}

\newcommand{\fix}[1]{{\color{red}{#1}}}

\newboolean{anonymous}
\setboolean{anonymous}{false}
\newcommand{\switch}[3]{\ifbool{#1}{#2}{\fix{#3}}}
\newcommand{\comment}[1]{}

\newdimen\origiwspc
\newdimen\origiwstr
\newcommand{\crush}[1]{{\newdimen\origiwspc \origiwspc=\fontdimen2\font \fontdimen2\font=1pt #1 \fontdimen2\font=\origiwspc}}

\begin{document}

\title{{Sharing FANCI Features: A Privacy Analysis \\ of Feature Extraction for DGA Detection}}
\switch{anonymous}
{\author{\IEEEauthorblockN{\emph{Anonymous Author(s)}}}}
{
\author{\IEEEauthorblockN{Benedikt Holmes}
\IEEEauthorblockA{
\textit{RWTH Aachen University}\\
Aachen, Germany \\
holmes@itsec.rwth-aachen.de}
\and
\IEEEauthorblockN{Arthur Drichel}
\IEEEauthorblockA{
\textit{RWTH Aachen University}\\
Aachen, Germany \\
drichel@itsec.rwth-aachen.de}
\and
\IEEEauthorblockN{Ulrike Meyer}
\IEEEauthorblockA{
\textit{RWTH Aachen University}\\
Aachen, Germany \\
meyer@itsec.rwth-aachen.de}
}
}

\maketitle

\copyrightnotice{\copyright\space Copyright held by the owner/author(s) 2021. This is the author's version of the work. It is posted here for your personal use. Not for redistribution. The definitive version was published in The Sixth International Conference on Cyber-Technologies and Cyber-Systems (CYBER 2021), https://www.thinkmind.org/index.php?view=article\&articleid=cyber\_2021\_1\_160\_80095}

\begin{abstract}
The goal of Domain Generation Algorithm (DGA) detection is to recognize infections with bot malware and is often done with help of Machine Learning approaches that classify non-resolving Domain Name System (DNS) traffic and are trained on possibly sensitive data.
In parallel, the rise of privacy research in the Machine Learning world leads to privacy-preserving measures that are tightly coupled with a deep learning model's architecture or training routine, while non deep learning approaches are commonly better suited for the application of privacy-enhancing methods outside the actual classification module.
In this work, we aim to measure the privacy capability of the feature extractor of feature-based DGA detector FANCI (Feature-based Automated Nxdomain Classification and Intelligence).
Our goal is to assess whether a data-rich adversary can learn an inverse mapping of FANCI's feature extractor and thereby reconstruct domain names from feature vectors.
Attack success would pose a privacy threat to sharing FANCI's feature representation, while the opposite would enable this representation to be shared without 
privacy concerns.
Using three real-world data sets, we train a recurrent Machine Learning model on the reconstruction task.
Our approaches result in poor reconstruction performance and we attempt to back our findings with a mathematical review of the feature extraction process.
We thus reckon that sharing FANCI's feature representation does not constitute a considerable privacy leakage.

\end{abstract}


\section{Introduction}
Machine Learning (ML) has had great success in solving advanced data-driven problems and its application also yields great performance for solving the Domain Generation Algorithm (DGA) classification problem.
Instead of using static IP-addresses or domain names, bots use DGAs to generate pseudo-random domain names and then query the Domain Name System (DNS) to obtain the IP address of their command and control server. The botnet herder knows the DGA generation scheme and is therefore able to register a subset of the generated domains, while the connection is now more difficult for the defender to block.
Most of the bot's queries result in non-existing domain (NXD) responses as only the domain names that are registered in advance are resolved to valid IP addresses.
ML classifiers can be trained to separate benign NXDs, e.g., caused by typos or misconfigured software, from DGA generated domains.
Thereby, DGA activities can be detected even before bots receive instructions from the herder.

For reasons such as the availability, diversity, or size of data, it is uncommon that ML models are trained solely on a large data set obtained from a single source. On the other hand, collecting or sharing sensitive data is a privacy-concern.
For ML-based DGA detection on NX traffic, the malicious training samples are publicly sourced, e.g., obtained from DGArchive \cite{plohmann_comprehensive_2016}, while samples of benign NXDs are often locally collected and can contain privacy-sensitive information as their disclosure may allow drawing conclusions about sensitive activity on the network, e.g., usage of a particular software or end-user browsing.
Deep Learning (DL) is designed to allow models to directly receive raw data as input and therefore privacy-preserving measures are often coupled with the training routine.
Non-DL approaches are commonly preceded by a feature extraction stage that performs a data transformation with the goal of reducing size of the data while increasing expressiveness by compressing data samples to finite and fixed length vectors. Whether such transformation can also yield a sufficiently abstract data representation able to hide sensitive information is our main research focus.

In this work, we thus take a step back from the advances in DL-based DGA detection and reconsider a simpler, feature-based DGA detection approach and evaluate its practicability towards privacy-preserving intelligence sharing:
FANCI (Feature-based Automated Nxdomain Classification and Intelligence) \cite{schuppen_fanci_2018} is the first feature-based classifier that achieves significant performance in DGA detection while only considering few hand-crafted features.
Complementing the research on its classification performance \cite{schuppen_fanci_2018}, we investigate whether FANCI's public feature extractor is prone to malicious inversion. More concretely, we ask whether knowledge of FANCI features threatens the disclosure of the original 
domain names as the latter could be reconstructable from features.
If the feature extraction process can be deemed inversion-resilient, then this allows the risk-free publication of sensitive NX data in the form of FANCI's feature representation and would thus enable to provide data-privacy in otherwise privacy-concerning sharing tasks, e.g.,
(1) collaborative learning approaches in which many parties join their data or
(2) classification outsourcing in which DGA detection is offered as a service.

Our approaches exhibit poor reconstruction performance even when provided with real-world data samples. Consequently, we believe that FANCI's feature extractor is hard to invert, which motivates low-risk publication of feature vector sets for aforementioned sharing scenarios.

The work is structured as follows: Sections \ref{sec:related_work} \& \ref{sec:preliminaries} elaborate on relevant related work and preliminaries such as FANCI and its feature extractor. Section \ref{sec:math_review} gives a mathematical review of the feature extraction process to assess the limitations of any reconstruction approach. Then, these insights are used to motivate the subsequent data-driven approach, detailed in Section \ref{sec:methodology} \& \ref{sec:data_recon}, in which a DL model is trained to learn a reconstruction mapping based on three large real-world NX data sets. These allow us to asses whether a reconstructor trained on one data set may perform well on another data set at test time. Results, quantified by a normalized edit distance, are presented in Section \ref{sec:results} followed by a discussion in Section \ref{sec:discussion}. Finally, Section \ref{sec:conclusion} concludes the paper with an outlook on future work.

\section{Related Work}\label{sec:related_work}

We briefly give an overview of DGA detection methodologies and position ourselves in the research area of ML privacy.

\subsection{DGA Detection}
A variety of different DGA detection techniques have been proposed in the past, which can broadly be divided into context-less~\cite{schuppen_fanci_2018,woodbridge_predicting_2016,yu_character_2018,saxe_expose_2017,drichel_analyzing_2020} and context-aware approaches~\cite{antonakakis_throwaway_2012,bilge_exposure_2014,grill_detecting_2015,yadav_winning_2012,schiavoni_phoenix_2014,shi_malicious_2018}.
Context-less approaches only use information that can be extracted from a single domain name to determine whether a domain name is benign or malicious while context-aware approaches use additional contextual information to improve classification performance.
Previous studies suggest that context-less approaches achieve state-of-the-art detection performance while being less resource intensive and less privacy invasive than context-aware approaches~\crush{\cite{schuppen_fanci_2018,yu_character_2018,woodbridge_predicting_2016}~\cite{drichel_analyzing_2020}}.

The context-less approaches can further be divided into feature-based classifiers such as random forests or support vector machines (e.g.,~\cite{schuppen_fanci_2018}), and feature-less classifiers such as recurrent, convolutional, or residual neural networks \crush{\cite{woodbridge_predicting_2016}~\cite{yu_character_2018}~\cite{drichel_analyzing_2020}.}
The former group uses domain knowledge to extract hand-crafted features from a single domain name prior to classification.
The latter group of approaches consists of DL classifiers that learn to extract valuable features on their own, yet require many training samples.

The main object under study is the context-less and feature-based DGA detector FANCI \cite{schuppen_fanci_2018} that comprises a feature extractor and implements a random forest classification module.

\subsection{Privacy in Machine Learning}
ML has become the main suspect of privacy research investigating threats and defenses regarding models' and training procedures' natural information leakage of the consumed sensitive training data (e.g., \cite{ateniese_hacking_2015, fredrikson_model-inversion_2015, shokri_membership-inference_2017, papernot_sok_2018, al_rubaie_ppml_2019, nasr_comprehensive_2019}).
The attack class that relates closest to our work is Model Inversion \crush{\cite{fredrikson_model-inversion_2015}~\cite{fredrikson_warfarin_2014}} which, for a given model output, iteratively searches for the best fit input candidate based on some likelihood maximization scheme, e.g., by misusing loss and gradient of a neural network model. In our work however, the object under study is not a probability-outputting classifier, but rather just a data transformation module.
While the general goal of our studied threat and that of Model Inversion are aligned (finding a suitable input for a given output), our work more closely matches the terminology of \cite{al_rubaie_ppml_2019}: There, the term \emph{reconstruction} specifies malicious inversions of the feature extraction stage with the goal to map features back to raw training data samples.

\section{Preliminaries}\label{sec:preliminaries}
This section briefly introduces FANCI's feature extractor and presents the concept of Sequence-to-Sequence learning, which we leverage as reconstruction tool later in the study.

\subsection{FANCI'S Feature extractor}
We utilize the most recent open-source implementation of FANCI's feature extractor \cite{fanci_implementation}, which extracts 15 structural, 8 linguistic, and 22 statistical features from domain names as listed in Table~\ref{tab:fanci_features}.
For some features, an according footnote highlights that the implementation deviates from the definition in the original paper \cite{schuppen_fanci_2018}, e.g., \texttt{\small contains\_ipv4\_addr} should also regard IPv6 addresses.
The feature extraction recognizes 39 unique characters (letters a-z, digits 0-9 and special characters dot, hyphen and underscore).
For our study, we flatten the representation of the feature vector: Feature \texttt{\small number\_of\_subdomains} is represented as a one-hot encoded vector and clips the number of sub-domains at value 4.
Although it is in fact just one feature, we keep the representation via four values. Similarly, we also view each entry of the one-, two-, and three-gram distribution vectors as single feature. Thereby, our feature count differs slightly from the one presented in the original work \cite{schuppen_fanci_2018} and we end up with 45-component feature vectors.

As marked in Table~\ref{tab:fanci_features}, many of FANCI's features are computed on the Dot-free public-Suffix-Free (DSF) part of the domain which excludes both dot characters and the public valid suffix, which is usually the Top-Level-Domain (TLD). The validity of a suffix is determined by checking against a predefined list that is included in the feature extractor.

For the rest of this work, we formally refer to the feature extractor as a function $E\colon S \rightarrow F$ mapping domains from $S$ to the feature space $F$, where $S$ is the set of strings over the 39-character alphabet with lengths up to 253.

\begin{table}[tb]
\caption{Features extracted for FANCI}
\centering
\resizebox{\columnwidth}{!}{%
\begin{tabular}{c|l|c|c|c}
Feature  & Name & Type & Choices & Normalized by \\
\hline & & & & \\[-2.0ex]
1        & length                                                  & integer  & 250           & 253 \\
2-5      & number\_of\_subdomains$^{\mathrm{a}}$                   & integer  & 1             & 1 \\
6        & subdomain\_lengths\_mean$^{\mathrm{a}}$                 & rational & 250           & length \\
7        & contains\_wwwdot                                        & binary   & 2             & 1 \\
8        & has\_valid\_tld                                         & binary   & 2             & 1 \\
9        & one\_char\_subdomains$^{\mathrm{a}}$                    & binary   & 2             & 1 \\
10       & prefix\_repetition                                      & binary   & 2             & 1 \\
11       & contains\_tld\_as\_infix$^{\mathrm{a}}$                 & binary   & 2             & 1 \\
12       & only\_digits\_subdomains$^{\mathrm{a}}$                 & binary   & 2             & 1 \\
13       & only\_hex\_subdomains\_ratio$^{\mathrm{a}}$             & rational & 250+1         & 1 \\
14       & underscore\_ratio$^{\mathrm{a}}$                        & rational & 250+1         & 1 \\
15       & contains\_ipv4\_addr$^{\mathrm{a},\mathrm{c}}$          & binary   & 2             & 1 \\
\hline & & & & \\[-2.0ex]
16       & contains\_digits$^{\mathrm{a}}$                         & binary   & 2             & 1 \\
17       & vowel\_ratio$^{\mathrm{a},\mathrm{b}}$                  & rational & 250+1         & 1 \\
18       & digit\_ratio$^{\mathrm{a},\mathrm{b}}$                  & rational & 250+1         & 1 \\
19       & char\_diversity$^{\mathrm{a},\mathrm{b},\mathrm{c}}$    & rational & 250           & 1 \\
20       & alphabet\_size$^{\mathrm{a},\mathrm{b}}$                & integer  & 38            & 38 \\
21       & ratio\_of\_repeated\_chars$^{\mathrm{a},\mathrm{b}}$    & rational & 38+1          & 1 \\
22       & consecutive\_consonant\_ratio$^{\mathrm{a},\mathrm{b}}$ & rational & 250+1         & 1 \\
23       & consecutive\_digits\_ratio$^{\mathrm{a},\mathrm{b}}$    & rational & 250+1         & 1 \\
\hline & & & & \\[-1.9ex]
         & for \emph{n} $\in\{1,2,3\}$:                            & & & \\
24,31,38 & \emph{n}-grams\_std$^{\mathrm{a},\mathrm{b}}$           & rational & 1             & \multirow{7}{*}{$\left( \text{\parbox{1.3cm}{features of \emph{n}-grams are \\ normalized by their respective max value}} \right)$} \\
25,32,39 & \emph{n}-grams\_median$^{\mathrm{a},\mathrm{b}}$        & rational & 250           & \\
26,33,40 & \emph{n}-grams\_mean$^{\mathrm{a},\mathrm{b}}$          & rational & 1             & \\
27,34,41 & \emph{n}-grams\_min$^{\mathrm{a},\mathrm{b}}$           & integer  & 250+1         & \\
28,35,42 & \emph{n}-grams\_max$^{\mathrm{a},\mathrm{b}}$           & integer  & 250           & \\
29,36,43 & \emph{n}-grams\_perc\_25$^{\mathrm{a},\mathrm{b}}$      & rational & 250           & \\
30,37,44 & \emph{n}-grams\_perc\_75$^{\mathrm{a},\mathrm{b}}$      & rational & 250           & \\
45       & shannon\_entropy$^{\mathrm{a},\mathrm{b}}$              & rational & $\approx$ 194 & $log_2(\text{\footnotesize alphabet\_size})$ \\
\hline
\multicolumn{2}{l}{\rule{0pt}{2ex}$^{\mathrm{a}}$Feature ignores public suffix.} &\multicolumn{2}{l}{$^{\mathrm{b}}$Feature ignores dots.} & \\
\multicolumn{5}{l}{\rule{0pt}{2ex}$^{\mathrm{c}}$Definition of feature in implementation deviates from original paper.} \\
\end{tabular}
}
\label{tab:fanci_features}
\end{table}

\subsection{Sequence-to-Sequence Learning}
Sequence-to-Sequence learning (Seq2Seq) encompasses encoder-decoder models that solve ML tasks related to mapping variable-length input sequences to variable-length output sequences \crush{\cite{cho_learning_2014}~\cite{sutskever_sequence_2014}}.
Usually, both the encoder and decoder of a Seq2Seq architecture utilize recurrent units to process the variable-length sequences and work together as follows: The encoder consumes and compresses the input sequence to a fixed-length state while the decoder is trained to create the target sequence from this state.
A common use case is machine language translation on token sequences (i.e., words or characters).
To train the model, bounds of the output sequences must be encoded in some fashion such that the token-wise decoding process begins with a start marker and can be stopped once the end marker is encountered or predicted.
At test time a sequence can be sampled from the decoder of a trained Seq2Seq model, i.e., beginning with the start marker, the model iteratively predicts the next character with the currently sampled prefix as prior. This sampling technique is commonly referred to as closed-loop, since the predicted characters are fed back into the model at each step.

\section{A Mathematical Review}\label{sec:math_review}
Here we view the plain mathematical definition of the feature extraction process as such and assess the invertability of the process.
The goal of inversion is to find a valid function $E^{-1}\colon F \rightarrow S$. Due to the feature extractor $E\colon S \rightarrow F$ not being bijective, function $E^{-1}$ can obviously only fulfill $E^{-1}(E(s)) = s$ for samples $s\in R$ of a certain subset $R \subset S$ for which we are concerned that it includes real-world NXDs.
Estimating $E^{-1}$ by sampling a complete look-up table would require iterating over all domains in $S$. In theory, the domain space is of size $|S|=\sum_{i=4}^{253}39^{i}\approx 3.55\cdot10^{402}$.
Size of the feature space $F$ can be estimated based on the multiplication of possible choices for each value in a feature vector (see Table~\ref{tab:fanci_features}). 
For this, we respect the specification's maximum domain length of 253 \cite{rfc} and choose the minimum length to be four.
For the rational-valued features the number of choices is determined by the size of the divisor or if the divisor is another feature, then we view the number of choices as the maximum possibilities for the dividend. Occasionally, the dividend is allowed to be zero, which is denoted by a ``+1'' in Table~\ref{tab:fanci_features}.
For the \texttt{\small entropy} feature we estimate the average amount of distinct values it can accommodate. For features that are dependent on others we view the amount of choices as fixed, or ``1'' in Table~\ref{tab:fanci_features}. Finally, we approximate that $|F| \approx 2.71\cdot10^{65}$.
Consequently, the feature extraction process performs a reduction of order of magnitude $1.311\cdot10^{337}$, i.e., there may on average exist $10^{337}$ pre-images for each feature vector. In the best case, where all pre-images are distributed equally among all images, inversion would thus be impossible.

\subsection{Inference of new Information}
It is possible to infer new information about the original domain sample $s\in S$ via the combination of existing features in $f=E(s)$ and thereby more accurately capture the number of possible pre-images.
The information is new in the sense that it is not previously held directly as a value in $f$.
Examples of inferable information are listed in Table~\ref{tab:inferable_information}.
Further, Shannon entropy is computed as weighted sum of character frequencies $H=\sum_{i=1}^{n} freq(x_i) \cdot log(freq(x_i))$ with restrictions $\sum_{i=1}^{n} x_i = p_1$ and $x_i \in \mathbb{N}^{\ge1} \forall i \in \{1,...,n\}$. The underlying character frequency distribution $freq(\cdot)$ over $n$=\texttt{\small alphabet\_size} unknown characters is uniquely determined based on $p_1$ and \texttt{\small entropy}.

\begin{table}[b]
\caption{New information inferable from other FANCI features}
\centering
\resizebox{\columnwidth}{!}{%
\begin{tabular}{c|p{3cm}|c}
\textbf{ID} & \multicolumn{1}{c|}{\textbf{New Information}} & \textbf{Inference Rule} \\
\hline\hline
\rule{0pt}{3ex}$p_1$ & Length of the DSF & $p_1 = \frac{\texttt{\scriptsize alphabet\_size}}{\texttt{\scriptsize char\_diversity}}$ \rule{0pt}{3ex} \\
\hline
\rule{0pt}{3ex}$p_2$ & Amount of sub-domains & $p_2 = \frac{p_1}{\texttt{\scriptsize subdomain\_lengths\_mean}}$ \rule{0pt}{3ex} \\
\hline
\rule{0pt}{3ex}$p_3$ & Length of the public suffix & $p_3 = \texttt{\scriptsize length} - (p_1 + p_2)$ \rule{0pt}{3ex} \\
\hline
\rule{0pt}{3ex}$p_4$ & Total digit occurrences& $p_4 = \texttt{\scriptsize digit\_ratio} \cdot p_1$ \rule{0pt}{3ex} \\
\hline
\rule{0pt}{3ex}$p_5$ & Total vowel occurrences & $p_5 = \texttt{\scriptsize vowel\_ratio} \cdot (p_1 - p_4)$ \rule{0pt}{3ex} \\
\hline
\rule{0pt}{3ex}$p_6$ & Occurrences other chars & $p_6 = p_1 - (p_4 + p_5)$ \\
\end{tabular}
}
\label{tab:inferable_information}
\end{table}

\subsection{Limitations}
Finding the unique solution for the discrete frequency distribution that matches the entropy feature may require enumerating all solution candidates. Due to the way the entropy is calculated, the number of solution candidates is given by the binomial coefficient $\binom{n-1}{k-1}$.
It is further possible to estimate the amount of unique digits ($u_d$), vowels ($u_v$) and other characters ($u_o$) by iterating over all valid allocations of bins in the frequency distribution to one of the three groups such that the sum of frequencies for each group's bins matches with the previously inferred total count (i.e., $p_4$, $p_5$ and $p_6$). There is not necessarily a unique solution to this.

With help of combinatorics we can specify a tighter bound on the number of possible pre-images per feature vector $f$.
First, dsf-struct$(f)$ captures the possibilities to structure the DSF in \eqref{eq:dsf_structure}: Given the inferred total occurrences of digits ($p_4$), vowels ($p_5$) and others ($p_6$), one can virtually choose $p_4$ character slots in the DSF of length $p_1$ and similarly $p_5$ slots of the remaining $p_1-p_4$. The final slots are for the third group. Then, there are $\binom{p_1-1}{p_2-1}$ possibilities to split the DSF into $p_2$ sub-domains by inserting the $p_2-1$ separating dots.

\begin{equation}
\text{dsf-struct}(f) = \binom{p_1}{p_4} \cdot \binom{p_1-p_4}{p_5} \cdot \binom{p_1-1}{p_2-1}
\label{eq:dsf_structure}
\end{equation}

Secondly, for a fixed valid setting of unique character counts $u = (u_d, u_v, u_o)$ we can estimate the following:
Possibilities for digit occurrences in the DSF is determined by the choices of an $u_d$-large subset of all digits and all permutations of each subset of length of the total occurrences of digits in the DSF, i.e., $p_4$. Same holds for unique count of vowels ($u_v$ \& $p_5$) and others ($u_o$ \& $p_6$). Since there does not have to be a unique solution for $u$, one needs to sum over the possible choices for $u$.
Similarly, we can thus capture the total amount of possibilities for the DSF's content by dsf-cont$(f)$ in \eqref{eq:dsf_content}:

\begin{equation}
\text{dsf-cont}(f) = \sum_{u} \binom{10}{u_d} \cdot \binom{5}{u_v} \cdot \binom{24}{u_o} \cdot {u_d}^{p_4} \cdot {u_v}^{p_5} \cdot {u_o}^{p_6}
\label{eq:dsf_content}
\end{equation}

Finally, the public suffix list used by the feature extractor fixes the amount of choices for the public suffix or TLD with known length to $t_f$ and in total this results in \eqref{eq:preimages}, which more reasonably models the number of possible pre-images.

\begin{equation}
\begin{gathered}
\text{pre-images}(f) = t_f \cdot \text{dsf-struct}(f) \cdot \text{dsf-cont}(f)
\end{gathered}
\label{eq:preimages}
\end{equation}

However, the number of pre-images still remains significantly large and improving this manual reconstruction approach ultimately fails due to the lack of more linguistic information: Even if a frequency distribution can be determined, then the allocation of characters to those frequencies remains undetermined as that information is not held in the feature vector itself. Clearly, the function is not bijective and it is impossible to distinguish between equally likely pre-images. Arguing, to which extent more useful information can be extracted or whether a different manual approach would be more beneficial is a complex matter, which is why we attempt to let a DL model learn a reconstruction mapping based on real-world data.

\section{Methodology}\label{sec:methodology}
In reality, neither the amount of valid pre-images is equally distributed among all possible feature vectors nor are all pre-images for one feature vector equally likely.
In fact, real-world examples for benign NXDs and their corresponding feature vectors will only make up a small fraction of the respective domain space $S$ and feature space $F$: Besides that some subspace of $S$ (and thereby also a subspace of $F$) is occupied by the malicious samples, benign NXDs that result from typographical errors may still exhibit linguistic characteristics that are of low-entropy. For feature vectors, we argue that there are semantically invalid combinations of features, e.g., \texttt{\small alphabet\_size} = 1 while both \texttt{\small vowel\_ratio} $>$ 0 and \texttt{\small digit\_ratio} $>$ 0. Subsequently, the feature extractor will only act on a restriction of the mapping $S \rightarrow F$ in reality.

True pre-image distributions and domain-feature relations are best captured by real-world NXD samples, and hence, we leverage such data sets two-fold:
(1) To train a DL model that may learn the distribution of the sample data and
(2) as ground truth to assess the reconstruction capability of the trained models. The rest of this section defines the methodology for the experiment, and the evaluation of a DL reconstructor.

\subsection{Attack Model}
The context in which the following experiment is conducted is defined by the following aspects:
(1) We assume an adversary that is interested in learning the real inputs to the FANCI feature extractor $E: S \rightarrow F$ for a foreign feature set $S' \subset S$ of a target, i.e., for any $E(s) = f\in F$, the adversary aims to find a corresponding $s'\in S$ such that $E(s') = f$ holds and some closeness $s'\approx s$ is satisfied (Note, that finding just any $s'\in S$ with $E(s')=f$ is trivial).
(2) The adversary is semi-honest, i.e., he reliably participates in any sharing scenario through which he acquires the foreign feature set.
(3) The feature extractor is public knowledge.
(4) We only consider the disclosure of benign NXDs as privacy critical.
(5) We assume feature sets are shared in the clear, hence no interaction with the target is required.
(6) We allow the adversary to be in possession of an arbitrary large data set $S'' \subset S$ of benign NXDs that does not intersect with the target's data, i.e.,  $S'' \cap S' = \emptyset$.
(7) We do not restrict the adversary's computational power that he may apply to his own data. Hence, we allow the adversary to train an ML model.

\subsection{Reconstruction Quantification}
We leverage existing members from the family of edit distances on the string space to compare pairs of original and reconstructed samples.
Due to the possible encounter of unequal lengths, the only suitable candidates are the
\emph{Levenshtein} \cite{levenshtein_1966} distance metric and its variant \emph{Damerau-Levenshtein}, which both compute a minimum-change distance via the number of character edit operations (substitutions, insertions or deletions) required to transform the one input string into the other. We use the latter of both metrics which additionally considers the transposition of adjacent characters as a single operation.
Further, we compute a normalized version for the metric by dividing the resulting minimum-change-distance by the length of the longest of both input strings. This division operation invalidates none of the metric's axioms.
Note, however, that the normalized metric is a ratio of edit-operations to string length and which can be interpreted as a lower bound on the percentage of misplaced characters in the longer input string.

Consequently, a metric value of zero indicates equality, while dissimilarity grows in parallel to larger metric values. 
Thereby, quantifying the closeness, as previously mentioned in the attack model, becomes possible. Note that the choice of any threshold $\varepsilon > |s-s'|$ indicating attack success is subjective, as the metric does not regard a semantic comparison.

\subsection{Benign Data Sets}
In the following, we briefly comment on the nature and origin of the real-life benign NXD data we use, that are sourced locally by distinct institutions in different countries.

\subsubsection{\textbf{University\textsubscript{A}}} 
RWTH Aachen University in Germany provided us with a record comprising approximately 26 million unique NXDs recorded in the month of September 2019 by their central DNS resolver.
This resolver handles academic and administrative networks, the university hospital as well as networks of student residences.

\subsubsection{\textbf{University\textsubscript{B}}} 
We obtained another data set comprising 8 million unique samples recorded between mid-May 2020 and mid-June 2020 at Masaryk University that is located in the Czech Republic.

\subsubsection{\textbf{Association}} 
CESNET is a 27-member association of Czech universities which develops and operates a national e-infrastructure for science, research, and education, including several university networks.
We obtained a partial quantity of a one-day recording on June 16\textsuperscript{th} 2020 containing approximately 362k unique samples.

We use the complete record of the Association and draw a random sub-sample in the size of the Association's record from each of the other two institutions' records.
Intersections with one another and with malicious samples drawn from the open source intelligence feed of DGArchive \cite{plohmann_comprehensive_2016} (up until September 1\textsuperscript{st} 2020) are removed from all records prior to sub-sampling.

\subsection{Evaluation Setup}
In the following experiment, we assess the reconstruction performance of trained DL reconstruction models via the above mentioned distance metrics. More concretely, we train a DL model for each one of the benign data sources and evaluate each of these models against all the data sources including the one on which the individual model was trained on. For each pair of training and evaluation sets we average each metric's scores over all samples. Thereby, we assess the models' capability to reconstruct domains from foreign feature sets.

\section{Data-Driven Reconstruction}\label{sec:data_recon}
The following describes the training setup for the Seq2Seq decoder which is trained on the task of domain sample reconstruction (i.e., learning an inverse mapping \mbox{$E^{-1}|_R\colon F \rightarrow S$} on subset $R$) using an attack set of benign NXDs and their corresponding feature vectors $\{(f_i, s_i)\}_{i=0}^{n}\subset F \times S$. 
This is a realistic scenario in any sharing use case where a party receiving a feature set may also be in possession of an own data set of benign NXDs.
Basically, we assume that the feature extractor is unknown to the model, and we let it learn the inverse mapping without any domain-specific assistance.

\subsection{Model Architecture}
To reconstruct a variable-length domain sample from a fixed-length feature vector, the decoder of a Seq2Seq model is utilized. All models share the same architecture whose design follows a related approach \cite{sutskever_sequence_2014}.
Beginning with two parallel sequences of two dense layers with 200 units each, this leaves opportunity for the model to manipulate the representation of the input feature vector before the two outputs are used as the initial states for the recurrent unit in the decoder.
For the recurrent unit, a single Long Short-Term Memory (LSTM) layer with 200 units is used.
Finally, the model ends with a dense layer of size 42 and a softmax activation to output a prediction vector over all relevant characters, which includes the 39 recognized domain characters plus start, end and empty markers used internally for the sequence encoding of domains. 
In total, the architecture comprises 301,642 trainable weights.

\subsection{Training Setup}
For a good balance between training time and model performance, we fix a batch size of 64 for our experiment.
Models are trained using the cross-entropy loss to penalize wrong character predictions. Being unaware of any unbalance-bias, a focal loss is used to dynamically down-weigh well-classified samples in the cross-entropy loss during training \cite{lin_focal_2018}.

Training data is prepared as follows: The test set is a random 20\% split of the total data. Another random 5\% split of the remaining training data is used as validation set.
All entries in a FANCI feature vector are in some finite bounded range of the non-negative rationals and are normalized to the range of $[0,1]$ by dividing each entry by the upper bound of its value range (see last column of Table~\ref{tab:fanci_features}). Domain names are encoded to character sequences with start and end markers.

Each model is allowed to train for at most 1000 epochs, while the training data is shuffled after each epoch and training is stopped early whenever 10 epochs without improvement of the validation loss are exceeded.
We follow the common methodology to train a Seq2Seq model and thus employ Teacher Forcing to train the decoder \cite{williams_teacher_forcing_1989}. This essentially sets the input of the decoder to the target sequence shifted by one time step (open loop) instead of feeding the decoder's outputs of previous time steps back into the model (closed loop).

\section{Results}\label{sec:results}
For each domain in an evaluation set, we sample a reconstructed domain from the trained reconstructor models using the normalized feature vector of the original domain as initial state input to the model. Averaged closed-loop reconstruction performance for all combinations of trained models and evaluation sets are given on the left side of Table~\ref{tab:seq2seq_results}.

\begin{table*}[tb]
\caption{Closed Loop Reconstruction Performance of Seq2Seq Reconstructor \& Feature Space Overlap}
\centering
\begin{tabular}{c|c||S|S||S||S[table-format = 5.0]|S||S|S}
\multicolumn{2}{c||}{\multirow{2}{*}{\textbf{Network Data Source}}}  & \multicolumn{2}{c||}{\multirow{2}{*}{\parbox{2.75cm}{\textbf{Averaged Reconstruction Performance}}}} & \multicolumn{5}{c}{\multirow{2}{*}{\textbf{Feature Space Overlap}}}\\

\multicolumn{2}{c||}{} & \multicolumn{2}{c||}{} & \multicolumn{5}{c}{}\\

\textbf{Training} & \textbf{Evaluation}               & \multicolumn{1}{c|}{\textbf{Dam-Leven.}} & \multicolumn{1}{c||}{\textbf{norm.}}          & \parbox{1.9cm}{\centering \textbf{\#Unique FV$^{\mathrm{a}}$ (Training Data)}} & \parbox{1.5cm}{\textbf{Training $\cap$ Evaluation}}                                              & \parbox{1.25cm}{\centering \textbf{\% of Eval Data}}           & \parbox{1.75cm}{\centering \textbf{\#Unique FV$^{\mathrm{a}}$ (All Data)}} & \parbox{1.3cm}{\centering \textbf{\% of Total Data}} \\
\hline
\rule{0pt}{2ex}                      & \textbf{University\textsubscript{A}} & 47.85  & 0.51  & {\multirow{3}{*}{288118}} & {-} & {-} & {\multirow{9}{*}{3462}} & {\multirow{3}{*}{10.3}} \\
\textbf{University\textsubscript{A}} & University\textsubscript{B}          & 23.22 & 0.72 & & 5789 & 32.9 &  & \\
                                     & Association                          & 20.61 & 0.66 & & 4809 & 12.1 &  & \\
\cline{1-7} \cline{9-9}
\rule{0pt}{2ex}                      & University\textsubscript{A}          & 72.94 & 0.75 & {\multirow{3}{*}{182786}} & 5789 & 11.5 &  & {\multirow{3}{*}{30.7}} \\
\textbf{University\textsubscript{B}} & \textbf{University\textsubscript{B}} & 15.00 & 0.53 & & {-} & {-} & \\
                                     & Association                          & 18.61 & 0.61 & & 12985 & 41.8 &  & \\
\cline{1-7} \cline{9-9}
\rule{0pt}{2ex}                      & University\textsubscript{A}          & 62.07 & 0.67 & {\multirow{3}{*}{169921}} & 4809 & 24.6 &  & {\multirow{3}{*}{22.9}} \\
\textbf{Association}                 & University\textsubscript{B}          & 20.01 & 0.65 & & 12985 & 30.9 &  & \\
                                     & \textbf{Association}                 & 13.66 & 0.46 & & {-} & {-} & \\
\hline
\multicolumn{4}{l}{$^{\mathrm{a}}$FV = Feature Vectors.}
\end{tabular}
\label{tab:seq2seq_results}
\end{table*}

\subsection{Baseline Reconstruction Performance}
Rows in Table~\ref{tab:seq2seq_results} with a highlighted evaluation set show the average Damerau-Levenshtein metric score for the case that the evaluation data is equal to all the data used to train, validate and test the model, and is to be interpreted as baseline reconstruction performance of a trained model.

Although the models achieve a small training and test loss, reconstruction performance is mediocre: For University\textsubscript{A}, University\textsubscript{B} and the Association we measure that on average respectively 47.85, 15.00 and 13.66 character edit operation separate each original domain and its reconstruction. The normalized version of the metric measures an average score for University\textsubscript{A} and University\textsubscript{B} that is just larger than $0.5$, i.e., on average at least 50\% of characters in each reconstruction are misplaced.
For the Association, we measure a slightly smaller average score of $0.45$. It seems that the models' baseline reconstruction performance is similarly bad on all data sets.

\subsection{Transferability}
The remaining lines in Table~\ref{tab:seq2seq_results} demonstrate the trained models' reconstruction performance on data from foreign networks, i.e., exactly the scenario which we describe in our attack model.
In all cases the reconstruction error is higher than in the baseline cases (score higher than $0.65$) with the worst performance ($0.75$) in the case where the model trained on data from University\textsubscript{B} is evaluated on that of University\textsubscript{A}.

\section{Discussion}\label{sec:discussion}
After re-consideration of the mathematical review of the feature extractor, it is plausible that the overall reconstruction performance is poor. After all, FANCI's feature extractor considers only very few features and thereby performs a compression of such extent which is tolerable for good classification performance but hinders good reconstruction quality.
In the rest of this section we continue to argue about a quantifiable proof that $E$ is not injective when restricted to the subspace of real-world benign NXDs and review the adversary's theoretical information gain for well-reconstructed domains.

\subsection{Feature Space Overlap}
We place our experiment in the scenario in which adversary and target NXD data are disjunct.
This does, however, not imply that the sets of feature vectors of each respective data set are also disjunct.
Therefore, we also quantify the overlap in feature space for the three data sets used in this study on the right side of Table~\ref{tab:seq2seq_results}.
First, it is important to note that although every data set contains approximately 362k unique samples, the amount of unique feature vectors is significantly lower which clearly indicates collisions in the feature space. Secondly, for every combination of two distinct NXD data sets we have an intersection of non-trivial size in the feature space, e.g., 11.5\% of University\textsubscript{B}'s data intersects with University\textsubscript{A}'s and 41.8\% of its data with that of the Association.

A large overlap in the feature space most certainly leads to a degraded reconstruction performance, as for the same feature vector the model may learn to reconstruct a domain different from the one the adversary wants to sample at test time. The worst-performing baseline and transferability reconstructions (training data of University\textsubscript{B}) coincides with the largest feature space overlap w.r.t. all data sets (see Table~\ref{tab:seq2seq_results}).

\subsection{Top 10\% Reconstructions}
The adversary has no clear way of estimating the confidence of a single reconstruction without ground truth unless he conducts an own analysis of which type of domains are reconstructed well using a second data set.
Hence, we also discuss what he could potentially learn from good reconstructions by taking a closer look at the best 10\% of all reconstructions for the transferability cases:
The average reconstruction performance for the top 10\% lies at $0.276$. Further, approximately 45-55\% of the top 10\% performers are occupied by IPv4 and IPv6 reverse DNS lookups and 20-35\% by spam-related or other DNS-related services, e.g., DNS blacklists.

We argue that the models perform so well in reconstructing these types of domains with (1) these domains' contents being well-structured, (2) sharing a large suffix, and (3) standing out by containing a lot of numerical characters. Hence, they occupy the sparse areas of the feature space around features such as a high \texttt{\small digit\_ratio}, low \texttt{\small  subdomains\_lengths\_mean} or a True value for features such as \texttt{\small only\_digits\_subdomains} \texttt{\small contains\_ipv4\_addr}. Further, these NXDs do not necessarily originate from user typos but rather from misconfigured software. This would also better explain the high occurrence of these types of domains in the data.

The question remains whether knowledge of reverse look-ups and spam-services is privacy-sensitive information and we claim the opposite. After all, these domains do not reveal any information about end-user browsing or sensitive tooling usage in the network from which the data was sourced.

\section{Conclusion and Future Work}\label{sec:conclusion}
In this study we analyzed the data privacy capabilities of the feature-based DGA detector FANCI. The main goal was to answer whether feature vectors of FANCI disclose any sensitive information about the original domain names.
We provide mathematical reasoning for the success likelihood for any best-case reconstruction attempt and demonstrate that a manual approach of inferring sensitive information from combination of features has its difficulties and most certainly has its limitations:
Reconstruction cannot be easily performed on the basis of a single feature vector.

Therefore, we chose to emulate the logical approach a data-rich adversary would take, namely training an ML model to learn a reconstruction mapping. To provide significance to our results, we make use of three large real-world NXD sets fortunately made available to us. Finally, we find reconstruction performance of the trained models to be worse than desired:
On average at least half of all character from a reconstructed domain are misplaced in the baseline cases. The models only perform best on foreign network's data for reverse lookups and other not-sensitive NXDs likely originating from misconfigured software. We find this to be the result of these domains sharing a large portion of the higher-level domains and occupying a special niche in the feature space.

Consequently, our experiment suggest that an ML model aiding in the attack cannot reliably reconstruct NXDs from foreign networks' FANCI feature vectors which would be, however, the main use case in an attack.

Due to its universality, our data-driven analysis approach can be used in the future to perform a similar privacy analysis on other feature extractors used for DGA detection. The general concept of the data-driven analysis approach can also be used for a privacy analysis of feature-based classifiers in other ML use cases.

\section*{Acknowledgments}
The authors would like to thank Masaryk University, CESNET and Jens Hektor from the IT Center of RWTH Aachen University for providing NXD data.
This project has received funding from the European Union's Horizon 2020 research and innovation programme under grant agreement No 833418. 
Simulations were performed with computing resources granted by RWTH Aachen University under project rwth0438.

\bibliographystyle{IEEEtran}
\balance
\bibliography{bibliography}

\end{document}